\documentclass[aps,prl,twocolumn,groupedaddress,showpacs]{revtex4}

\newcommand{\mubold}         {\mbox{\boldmath$\mu$}}
\newcommand{\Omegabold}         {\mbox{\boldmath$\Omega$}}
\def\lap{\hbox{${_{\displaystyle<}\atop^{\displaystyle\sim}}$}} 
\def\gap{\hbox{${_{\displaystyle>}\atop^{\displaystyle\sim}}$}}

\usepackage{graphicx}

\begin{document}


\title{TeV $\mu$ Neutrinos from Young Neutron Stars} 


\author{Bennett Link}
\affiliation{Montana State University, Department of Physics, Bozeman MT
59717, USA}
\email[]{link@physics.montana.edu}
\altaffiliation{Also Department of Physics ``Enrico Fermi'', University of
Pisa, Italy}
\author{Fiorella Burgio}
\affiliation{INFN Sezione di Catania, Via S. Sofia 64, I-95123 
Catania, Italy}
\email[]{fiorella.burgio@ct.infn.it}


\date{\today}

\begin{abstract}

Neutron stars are efficient accelerators for bringing charges up to
relativistic energies. We show that if positive ions are accelerated
to $\sim 1$ PeV near the surface of a young neutron star
($t_{\rm age}\lap 10^5$ yr), protons interacting with the star's
radiation field will produce beamed $\mu$ neutrinos with energies of
$\sim 50$ TeV that could produce the brightest neutrino
sources at these energies yet proposed. These neutrinos would be coincident
with the radio beam, so that if the star is detected as a radio
pulsar, the neutrino beam will sweep the Earth; the star would be a
``neutrino pulsar''. Looking for $\nu_\mu$
emission from young neutron stars will provide a valuable probe of the
energetics of the neutron star magnetosphere. 

\end{abstract}

\pacs{97.60.Gb,
      94.30.-d,
      95.85.Ry
}

\bigskip

\maketitle


Pulsar emission is interpreted as a consequence of the acceleration of
charges to relativistic energies somewhere in the neutron star's
magnetosphere. As the charges (positive ions or electrons) move along
the curved magnetic field lines, they produce curvature radiation
which in turn produces a pair cascade and beamed radiation (see
\cite{mestel98} for a review).  A long-standing and unsolved problem
in neutron star magnetospheric physics is where charges obtain most of
their energy. Near the stellar surface, charges are constrained to
move along magnetic field lines. In the presence of the plasma, the
field lines are approximate equipotentials; the electric and 
magnetic fields satisfy ${\mathbf E}\cdot{\mathbf B}\simeq 0$ for a
quasistatic magnetosphere. In most theories of pulsar emission,
charge acceleration occurs in a charge depleted polar region very near
the stellar surface, sustained by a deviation of the magnetosphere
from corotation with the star
\cite{rs75,as79,hm98}. An alternative is the ``outer gap'' model
\cite{chr86} in which the acceleration site can be far from the
stellar surface. To determine exactly where and how the charges are
accelerated would require a detailed description of the neutron star
magnetosphere, which we currently lack. Here we show that young
neutron stars might be among the brightest sources of muon neutrinos yet
proposed, and so might be the first sources to be detected by planned
or operating neutrino telescopes such as AMANDA-II and IceCube
\cite{halzen03}, and ANTARES, NEMO and NESTOR
\cite{carr03}. The detection of young neutron
stars as strong neutrino sources would be a fascinating breakthrough in
its own right, and would indicate that protons (or heavier ions) are
accelerated to high energies close to the star, thus providing an
important probe of the energetics and physical conditions that prevail
in the poorly understood magnetosphere of a neutron star. 

If the star's magnetic moment $\mubold$ and angular velocity
$\Omegabold$ satisfy $\mubold\cdot\Omegabold<0$ (as one would expect
for half of neutron stars), positive ions will be accelerated to
infinity \cite{gj69}.  If the neutron star is young, its surface will emit
brightly in soft x-rays, and the protons in accelerated nuclei will
scatter with this radiation field. If the protons are sufficiently
energetic, they will exceed the threshold for photomeson production
through the $\Delta$ resonance (the $\Delta^+$ is an excited state of
the proton, with a mass of $1232$ MeV). The $\Delta^+$ quickly decays
to a $\pi^+$, and muon neutrinos are produced through the following
channels:
\begin{equation}
p\gamma\rightarrow \Delta^+\rightarrow 
n\pi^+
\rightarrow 
n\nu_\mu\mu^+
\rightarrow
n e^+ \nu_e\bar{\nu}_\mu\nu_\mu.
\end{equation}
This idea for $\nu_\mu$ production was explored by \cite{zdmwh03} in
the context of magnetars. In this paper we explore a similar idea in a
different physical regime, that of young, hot neutron stars. We show
that young neutron stars could be strong sources of muon neutrinos,
with energies $\sim 50$ TeV, and with fluxes observable by large-area
neutrino observatories.

{\bf Energetics.}
Goldreich \& Julian \cite{gj69} showed that the potential drop {\em
across} the field lines of a pulsar from the magnetic pole to the last
field line that opens to infinity is of magnitude
$\Delta\Phi = \mu\Omega^2/2c^2$, 
where $\mu=BR^3$ is the stellar magnetic moment, $B$ is the strength
of the dipole component of the field at the magnetic poles and $R$ is
the stellar radius.  If charges in rapidly rotating pulsars with
unexceptional fields ($B\sim 10^{12}$ G) could be accelerated by this
potential drop, the corresponding energy would be huge (neglecting
radiation losses, but see below)
\begin{equation}
\epsilon_\Phi = eZ\Delta\Phi = 7 Z B_{12} p_{\rm ms}^{-2}
\mbox{ EeV},
\label{emax}
\end{equation}
per ion, where $e$ is the electronic charge, $Z$ is the charge
number of the ion, $B_{12}\equiv B/10^{12}$, $p_{\rm ms}$ is the
spin period in milliseconds and $R\simeq 10^6$ cm is the stellar
radius. Near the star, charges are constrained to follow the magnetic
field lines. For a charge to be accelerated, ${\mathbf
E}\cdot{\mathbf B}\ne 0$ is required. Moreover, there must be at least
some acceleration near the stellar surface, to supply the charges that
support the magnetosphere. {\em As a conjecture to be tested by neutrino
observations, we shall assume that a strong, accelerating field exists
near the stellar surface}. In this case, radiation losses will
determine whether or not nuclei can attain energies of the magnitude
given by eq. [\ref{emax}]. The power dissipated by a charge $q$ of
energy $\epsilon$ moving along a field line of radius of curvature
$\rho_c$ is $d\epsilon/dt=(2/3)(q^2c/\rho_c^2)(\epsilon/mc^2)^4$. The
acceleration time scale is $\sim \rho_c/c$, giving $d\epsilon/dt\simeq
\epsilon c/\rho_c$. Near a pole of a magnetic dipole, $\rho_c\gap R$. 
Radiation losses limit the acceleration {\em per nucleon} of a nucleus of
mass number $A$ to be $\epsilon_{\rm max}\simeq 20 (A/Z^2)^{1/3}$ PeV.
Hereafter we assume $Z/A\simeq 1/2$.

Now we estimate the photomeson production threshold to compare with
the radiation loss limit. 
The threshold condition for a proton to reach the $\Delta$ resonance
is that the proton and photon energies, $\epsilon_p$ and
$\epsilon_\gamma$, respectively, satisfy
\begin{equation}
\epsilon_p\epsilon_\gamma\ge 0.3\ {\rm GeV}^2 f_g, \quad\quad 
f_g\equiv (1-\cos\theta_{p\gamma})^{-1},
\end{equation}
where $\theta_{p\gamma}$
is the incidence angle between the proton and the photon in the lab
frame. This condition is independent of whether the proton is free or
bound in a nucleus; in either case, photomeson production occurs
in nearly the same way. In the rest frame of the nucleus, the photons have an
energy of about 300 MeV. This energy is much larger than the binding
energy per nucleon, so nuclear binding has little effect on the
conversion of protons. 

Young neutron stars ($t_{\rm age}<10^5$ yr) typically have
temperatures of $T_\infty\simeq$ 0.1 keV ($=1.2\times 10^6$
K). Typical photon energies near the surface are $\epsilon_\gamma=2.8
kT_\infty(1+z_g)\sim 0.4$ keV, where $z_g\simeq 0.4$ is the
gravitational redshift. The proton threshold energy for the delta
resonance is then $\epsilon_{p,{\rm th}}\simeq T_{\rm 0.1keV}^{-1}f_g$
PeV, where $T_{\rm 0.1keV}\equiv (kT_\infty/0.1\mbox{ keV})$. As we
show below, the conversion, if it happens, must occur near the stellar
surface. Hence, protons scatter with photons with a lab frame
scattering angle of $\theta_{p\gamma}\simeq 90^\circ$, giving
$f_g\simeq 1$. (We neglect the effects of gravitational light
bending.)  In comparing the threshold energy $\epsilon_{p,{\rm th}}$
to the radiation loss limit $\epsilon_{\rm max}$, it appears that
radiation losses will not preclude protons from reaching the $\Delta$
resonance, even for heavy nuclei.

Is the accelerating potential strong enough to bring protons to the
$\Delta$ resonance? The maximum energy per proton is
$\epsilon_\Phi/A$. Since the protons are scattering with radiation
that is being radiated isotropically from the star (we ignore the
possibility of temperature variations across the surface),
$\theta_{p\gamma}$ is $\le 90^\circ$, and so $f_g> 1$. For the
$\Delta$ resonance to be reached, we require that
\begin{equation}
B_{12} p_{\rm ms}^{-2} T_{\rm 0.1keV} \ge 3\times
10^{-4}. 
\label{threshold}
\end{equation}
This condition, provided that
$\mubold\cdot\Omegabold<0$, represents the most optimistic condition
for neutrino production since we have assumed that the full potential
$\Delta\Phi$ is available for accelerating ions.
Eq. [\ref{threshold}] is satisfied in many young pulsars, though not
all. Magnetars, which have typical fields of $10^{15}$ G, spin rates
of $4-8$ s and $T_{\rm 0.1keV}\simeq 4-6$, are borderline neutrino
emitters according to eq. [\ref{threshold}], as found by
\cite{zdmwh03}. In these estimates, we have neglected the possible
quenching of the accelerating field by pair cascades produced through
inverse-Compton scattering of the protons with the radiation
field; numerical simulations of this effect indicate that the
quenching will be negligible, even for the high surface temperatures
we are considering \cite{hm02}. Failure to detect neutrinos from
pulsars, however, could mean that the field is strongly quenched.

{\bf Conversion Probability.}  For a significant neutrino flux to
exist, the conversion probability for $p\gamma\rightarrow\Delta^+$
must be sufficiently high. The value of $f_g$ averaged over the
surface increases rapidly with distance from the star, hence, the conversion
must occur near the stellar surface or the threshold becomes
unattainable. We suppose that the $\Delta$ resonance is reached before
an altitude at which photons with $f_g\le 2$, become unavailable. The
photons with the lowest values of $f_g$ originate from the stellar
limb.  To crudely estimate $f_g$, we consider
$\theta_{p\gamma}$ to be determined by the angle between the center of
the star and the stellar horizon as ``seen'' by a proton at $r>R$,
measured from the stellar center. With this definition, $f_g=2$
corresponds to $r=1.2R$.  Some models of charge acceleration in pulsar
magnetospheres find that significantly charge-starved regions can
exist just above the stellar surface ($\sim 10^4$ cm), so that the
charges receive most of their acceleration there
\cite{rs75,as79,hm98}, well below the altitude at which $f_g$ begins
to exceed 2.

The mean free path for conversion at radius $r$ is
$l(r)=(n_\gamma\sigma_{p\gamma})^{-1}$, where $n_\gamma$ is the number
density of the radiation field at radius $r$ and
$\sigma_{p{\gamma}}\simeq 5\times 10^{-28}$ cm$^2$ is the cross
section for $\Delta^+$ production. Assuming isotropic radiation from
the stellar surface, the photon density there is $n_\gamma(R)=(a/2.8
k)([1+z_g]T_\infty)^3\simeq 9\times 10^{19}T_{\rm 0.1keV}^3$
cm$^{-3}$, where $a$ is the radiation density constant, and
$n_\gamma(r)=n_\gamma(R)(R/r)^2$. The mean free path at the stellar
surface is $l(R)\simeq 2\times 10^7 T_{\rm 0.1keV}^{-3}$ cm. The
probability $P$ that a proton has {\em not} been converted changes
with $r$ as $dP/P = -dr/l(r)$.  By the time a proton starting at the
surface has reached radius $r$, the probability of conversion is
$P_{\rm conv}(r)=1-P(r)$. Requiring conversion in the range $R\le r\le
1.2R$ gives $P_{\rm conv}\simeq 0.02 T_{\rm 0.1keV}^3$. For the
surface temperatures we are interested in ($T_{\rm 0.1keV}\simeq 1$),
the radiation field is quite optically thin to the conversion
process. Hence we can expect the conversion of only one proton per
nucleus, if there is a conversion at all. Because the energy imparted
to the $\Delta^+$ is much larger than the binding energy per nucleon,
the $\Delta^+$ will be ejected from the nucleus.  The pulsar emission
mechanism will be essentially unaffected, since so few nuclei are
affected. The probability of photodisintegration is negligible.

{\bf The Neutrino Energy.}  The average fraction of energy transferred
from the proton to the pion is $\sim 0.2$, or 200 $T_{\rm
0.1keV}^{-1}$ TeV, corresponding to a Lorentz factor of $\gamma_\pi =
1.4\times 10^6 T_{\rm 0.1keV}^{-1}$. The pion decays in a time $\tau_d
= \tau_\pi\gamma_\pi$, where $\tau_\pi=2.6\times 10^{-8}$ s, and thus
moves to $r\sim 10^3R$ before decaying. The pion is in the densest
part of the radiation field for only a short time, of order $R/c\simeq
3\times 10^{-5}$ s, and suffers negligible energy loss to
inverse-Compton scattering. When the pion decays, it gives 
one-fourth of its energy to the muon neutrino, the rest going to the
other three leptons. The resulting neutrino energy is then
\begin{equation}
\epsilon_{\nu_\mu} \simeq 50\ T_{\rm 0.1keV}^{-1}\mbox{ TeV}.
\label{enu}
\end{equation}

{\bf The Neutrino Flux.} The accelerated protons are far more
energetic than the radiation field with which they are
interacting. Any pions produced through the $\Delta$ resonance, and
hence, any muon neutrinos, will be moving in nearly the same direction
as the proton was when it was converted. We thus assume that the star
produces a neutrino beam that is nearly coincident with the radio
beam, that is, if the star is detected as a radio pulsar, the neutrino
beam will sweep the Earth. {\em The radio pulsar will also be a
``neutrino pulsar''.}

For a quasistatic magnetosphere, the charge density near the stellar
surface is $\rho_q \simeq eZn_0\simeq B/pc$ (cgs), where $n_0$
is the Goldreich-Julian number density of ions \cite{gj69}. To
get acceleration of charges, there must be a charge depleted gap
somewhere above the star. We parameterize the density in the gap as
$f_d n_0$, where $f_d<1$ is a model-dependent and uncertain depletion
factor. The depletion factor should be regarded as a free parameter,
though even modest depletion ($f_d\simeq 1/2$) could give enormous
accelerating fields, and a significant fraction of $\epsilon_\Phi$
could be reached \cite{rs75,hm98}. We see the radio beam for only a
fraction $f_b$ of the pulse period, the {\em duty cycle}. Typically,
$f_b\simeq 0.1-0.3$ for younger pulsars.  We take the duty cycle of
the neutrino beam to be $f_b$. The phase-averaged neutrino flux at
Earth resulting from the acceleration of positive ions, at a distance
$d$ from the source, will be
\begin{equation}
\phi_\nu\simeq c f_b f_d n_0 \left(\frac{R}{d}\right)^2
P_{\rm conv} 
\label{nuflux}
\end{equation}
Large-area neutrino detectors use the Earth as a
medium for conversion of a muon neutrino to a muon, which then
produces Cerenkov light in the detector. The conversion probability in
the Earth is $P_{\nu_\mu\rightarrow\mu}\simeq 1.3\times
10^{-6}(\epsilon_{\nu_\mu}/\mbox{1 TeV})$ where $\epsilon_{\nu_\mu}$
is the energy of the incident neutrino \cite{hh02}. Combining
eqs. [\ref{enu}] and [\ref{nuflux}] gives a muon event rate of 
\begin{eqnarray}
\frac{dN}{dAdt} &= &\phi_{\nu}P_{\nu_\mu\rightarrow\mu} \nonumber \\
& & \hspace*{-1.5cm} \simeq 10^5 Z^{-1}f_b f_d B_{12}p_{\rm ms}^{-1}d_{\rm
kpc}^{-2} T_{\rm 0.1keV}^2 \mbox{ km$^{-2}$
yr$^{-1}$}. 
\label{rate}
\end{eqnarray}
As an example, if $T_{\rm 0.1keV}=1$, $f_b=0.1$, $f_d=0.1$, $p_{\rm ms}=10$,
$d_{\rm kpc}=3$, $Z=2$, the muon flux would be $\sim 6$
km$^{-2}$ yr$^{-1}$ at a neutrino energy of $\simeq 50$ TeV. The
rate given by eq. [\ref{rate}] is model independent and quite general,
{\em provided that young neutron stars succeed in accelerating
positive ions to the $\Delta$ resonance.} 

{\bf Detectability}. 
For a neutron star to be a detectable neutrino pulsar, it must
satisfy eq. [4], and be sufficiently close to give a strong flux.
For a neutron star to be a detectable neutrino
pulsar, it must satisfy $B_{12}p_{\rm ms}^{-2} T_{\rm 0.1keV} >
3\times 10^{-4}$ (eq. [\ref{threshold}]), and be sufficiently close as
to give a strong flux. Potentially interesting sources and 
upper limits for their muon event rates are given in Table 1. We have
optimistically included SN1987a in the Table, in case this supernova
remnant contains a neutron star. We have estimated the background
above 50 TeV due to atmospheric cosmic ray events using the flux given
by \cite{gqrs98}; for detectors such as AMANDA-II and ANTARES, with
angular resolutions of $\sim 1^\circ$, the atmospheric background
above 50 TeV contributed over several years will be so low that the
detection of a {\em single muon} would represent a statistically
significant neutrino detection.

\begin{table}
\begin{ruledtabular}
\begin{tabular}{llllllll}
Source & $d_{\rm kpc}$ & age & $p_{\rm ms}$ & $B_{12}$ & $T_{\rm 0.1keV}$
& $f_b$ & $dN/dAdt$ \\
 & & yr & & & & & km$^{-2}$ yr$^{-1}$   \\
\colrule
Crab & 2 & $10^3$ & 33 & 3.8 & $\le 1.7$ \cite{weisskopf_etal04} & 0.14 & 1200 \\
Vela & 0.29 & $10^{4.2}$ & 89 & 3.4 & 0.6 \cite{pavlov_etal01} & 0.05 & 800 \\
J0205+64 & 3.2 & $10^{2.9}$ & 65 & 3.8 & $\le 0.9$ \cite{shm02}& 0.04 & 20 \\
B1509-58 & 4.4 & $10^{3.2}$ & 151 & 15 & 1? & 0.26 & 130 \\ 
B1706-44 & 1.8 & $10^{4.3}$ & 102 & 3.1 & 1? & 0.13 & 120 \\
B1823-13 & 4.1 & $10^{4.3}$ & 101 & 2.8 & 1? & 0.34 & 60 \\
Cass A & 3.5 \cite{pst04} & 300 & 10? & 1? & 4 \cite{pst04} & 0.1? & 1300 \\
SN 1987a & 50 & 17 & 1? & 1? & 4? & 0.1? &  60 \\
\end{tabular}
\end{ruledtabular}

\caption{Estimated $\mu$ Fluxes at Earth.
Magnetic fields were estimated from the vacuum dipole prediction:
$B=(3.2\times 10^{39}p\dot{p})^{1/2}$ G. Numbers followed by question
marks indicate guesses. We took $f_d=1$ and $Z=1$ so these rates
are upper limits, and must be scaled by a factor $f_dZ^{-1}$. Radio
pulsar spin parameters were taken from
\cite{psr_survey}. Temperatures and limits on temperatures were taken
from the references indicated. The temperature upper limits on the Crab
and J0205+64 were used.}

\end{table}

{\bf Discussion}. Results of 607 d of data from AMANDA-II are now available
\cite{ackermann_etal04,ahrens_etal04}. 
Cass A and the Crab pulsar were not detected as neutrino sources. Cass
A has not been observed as a pulsar, and so would not be expected to
be a source in this model. The Crab pulsar would be a strong source,
but only if $\mubold\cdot\Omegabold<0$. If the Crab is a neutrino
pulsar, and the temperature is close to the observational upper limit,
lack of detection by AMANDA-II requires $f_d\lap 0.01$ ($Z=2$), which
seems improbably low in the face of current models of charge depleted
gaps. It seems likely that the Crab is simply not a neutrino pulsar,
perhaps because $\mubold\cdot\Omegabold>0$. (The threshold condition
eq. [\ref{threshold}] would be easily met even for much lower
temperatures than the upper limit given in Table 1).

Should AMANDA-II have seen neutrinos from other pulsars? There are now
64 known pulsars that are within 10 kpc of Earth and younger than
$10^5$ yr \cite{psr_survey}, detected mostly by the Parkes Radiopulsar
Survey. Any of these is a potential neutrino pulsar. Because the
Parkes telescopes are located in the southern hemisphere, only 13 of
the 64 radio pulsars are in the northern sky, the part visible to
AMANDA-II. To make an estimate of the probability of detection of a
random neutrino pulsar by AMANDA-II, suppose that AMANDA-II were
located at the North Pole, observing southern-sky sources. We can then
use a sample of the 51 pulsars from the southern sky to assemble a
probability distribution for the rates. We took $T_{\rm 0.1keV}=1$,
$f_d=0.5$ and $Z=2$ in all cases, $f_b$ in the cases that it has been
measured, and $f_b=0.1$ in all other cases. Using these uncertain, but
very conservative numbers, we find only $\sim 1$ pulsar that satisfies
eq. [\ref{threshold}] and $\mubold\cdot\Omegabold<0$, so it is not
surprising that AMANDA-II has not yet seen a neutrino
pulsar. Given the small number of pulsars known in the
northern sky, our model cannot be statistically constrained by
AMANDA-II. AMANDA-II's successor, IceCube, with ten times the
effective area, might reveal neutrino pulsars. Moreover, in the
southern sky there are a number of promising candidates that might be
detected by ANTARES (expected to go into operation during 2005), but
that are inaccessible to AMANDA-II; these include Vela, B1509-58,
B1706-44, B1823-13 and {\em possibly} SN 1987a. There are other
candidates in the southern sky, none with measured temperatures, but
assuming $T_{\rm 0.1keV}\ge 1$, there are nine known pulsars that
satisfy eq. [\ref{threshold}], are within a distance of 5 kpc and are
younger than $10^5$ yr and therefore likely to have $T_{\rm
0.1keV}\sim 1$. These give typical muon event rates of $\sim
f_dZ^{-1}(20-130)$ km$^{-2}$ yr$^{-1}$; three of these candidates are
listed in Table 1.

The most promising candidate source in the southern sky is the Vela
pulsar, which if detected by ANTARES, would give an event rate of
$~80 Z^{-1}f_d$ yr$^{-1}$. For $Z=2$, the source would be
detectable over a year of observing time even if the acceleration
region is charge depleted to $\sim 3$\% of the Goldreich-Julian
density. For modest depletion ($f_d\sim 1/2$), Vela could be
detected in under three weeks. If we are unfortunate, and Vela has
$\mubold\cdot\Omegabold>0$, it will not be a neutrino pulsar, and we
must look for weaker sources.  For the other candidates to be
detectable by ANTARES, B1508-58 and B1706-44 might be
detected in about four months for $f_d\simeq 1/2$ and $Z=2$. After
the first year of data from ANTARES, statistical constraints will be
possible even if there are no detections, since the the properties of
pulsars in the southern sky are well understood. No detections would
mean that photomeson production is negligible or nonexistent, providing
an interesting constraint on the energetics of the magnetosphere. A
detection would allow a determination of $Z^{-1}f_d$, and hence, the
charge density in the acceleration region. We stress that the rates
appearing in Table 1 are upper limits; they could be significantly
reduced by the factor $Z^{-1}f_d$. 

We conclude by pointing out that neutrino pulsars could be among the
strongest sources of high-energy muon neutrinos in the sky, and
therefore, might be the first sources to be detected. (See
\cite{lm00,bbm05} for a review of other models). We will present the
energy spectrum of neutrino pulsars in a forthcoming publication.

\begin{acknowledgments}

We thank L. Bildsten, C. Distefano, C. M. Riedel, S. N. Shore,
C. Thompson and I. Wasserman for valuable discussions. BL thanks the
INFN at the Universities of Pisa and Catania. This
work was supported by the National Science Foundation under Grant
No. AST-0098728.

\vspace*{-5pt}

\end{acknowledgments}

\vspace*{-15pt}


\bibliography{references}

\end{document}